\documentclass[12pt]{article}
\usepackage{a4}
\begin{document}
\title{First order phase transition  and corrections to
  its parameters in the O(N) - model}
\author{
{\sc M. Bordag}\thanks{e-mail: Michael.Bordag@itp.uni-leipzig.de} \\
\small  University of Leipzig, Institute for Theoretical Physics\\
\small  Augustusplatz 10/11, 04109 Leipzig, Germany\\
\small and\\
{\sc V. Skalozub}\thanks{e-mail: Skalozub@ff.dsu.dp.ua}\\
\small  Dniepropetrovsk National University, 49050 Dniepropetrovsk, Ukraine}
\maketitle
\begin{abstract}
The temperature phase transition in the $N$-component scalar field
theory with spontaneous symmetry breaking is investigated using the
method combining the second Legendre transform and with the
consideration of gap equations in the extrema of the free
energy. After resummation of all super daisy graphs an effective
expansion parameter, $(1/2N)^{1/3}$, appears near $T_c$ for large $N$.
The perturbation theory in this parameter accounting consistently for
the graphs beyond the super daisies is developed.  A certain class of
such diagrams dominant in $1/N$ is calculated
perturbatively. Corrections to the characteristics of the phase
transition due to these contributions are obtained and turn out to be
next-to-leading order as compared to the values derived on the super
daisy level and do not alter the type of the phase transition which is
weakly first-order.  In the limit $N$ goes to infinity the phase
transition becomes second order.  A comparison with other approaches
is done.
\end{abstract}

\section*{1. Introduction}
Investigations of the temperature phase transition in the
$N$-component scalar field theory ($O(N)$-model) have a long history
and were carried out by either perturbative or non perturbative
methods. This model enters as an important part unified field theories
and serves to supply masses to fermion and gauge fields via the
mechanism of the spontaneous symmetry breaking. A general believe
about the type of the symmetry restoration phase transition is that it
is of second order for any values of $N$ (see, for instance, the text
books \cite{Zinn}-\cite{Kapusta}). This conclusion results mainly from
non perturbative analytic and numeric methods
\cite{Tetradis}-\cite{Fodor}. In opposite, a first order phase
transition was observed in most perturbative approaches
\cite{Takahashi}-\cite{pl}. An analysis of the sources of this
discrepancy was done in different places, in particular, in our
previous papers \cite{Bordag1}-\cite{pl} devoted to the investigation
of the phase transition in the $O(N)$-model in perturbation theory
(PT). Therein a new method has been developed which combines the
second Legendre transform with consideration of the gap equations in
the extrema of the free energy functional. This allows for
considerable simplification of calculations and for analytic results.
The main of them is the discovery in the so-called super daisy
approximation (SDA) of an effective expansion parameter $\epsilon =
\frac{1}{N^{1/3}}$ near the phase transition temperature $T_c$. All
quantities (the particle masses, the temperatures $T_+, T_-$) are
expandable in this parameter. The phase transition was found to be
weakly first order converting into a second order one in the limit $N
\to \infty$. The existence of this small parameter improves the status
of the resummed perturbative approach making it as reliable as any
perturbative calculation in quantum field theory. For comparison we
note that formerly there had been two problems with the perturbative
calculations near the phase transition. First, unwanted imaginary
parts had been observed. Using functional methods we showed in
\cite{Bordag1,pl} that they disappear after re-summing the super daisy
graphs. The second problem was that near $T_c$ the masses become small
(although not zero) compensating the smallness of the coupling
constant hence making the effective expansion parameter of order one.

In the present paper we construct the PT in the effective expansion
parameter $\epsilon = \frac{1}{N^{1/3}} $ near $T_c$ for the
$O(N)$-model at large $N$. It uses as input parameters the values
obtained in the SDA. As an application we calculate corrections to the
characteristics of the phase transition near $T_c$ which follow from
taking into account all BSDA graphs in order $\frac{1}{N}$.  Since
the masses of particles calculated in the SDA are temperature
dependent, we consider in detail the renormalization at finite
temperature of the graphs investigated. It will be shown that the
counter terms renormalizing these graphs at zero temperature are
sufficient to carry out renormalizations at finite temperature.

The paper is organized as follows. In sect.2 we adduce the necessary
information on the second Legendre transform and formulate the BSDA PT
at temperatures $T \sim T_c$.  In the next section we calculate the
contribution to the free energy of the graphs called the ``bubble chains''
having a next-next-to-leading order in $\frac{1}{N}$. In sect.4 the
renormalization is discussed. The corrections to the masses and other
parameters near $T_c$ are calculated in sect.5. The last section is
devoted to discussion.

\section*{2. Perturbation theory beyond super daisy approximation}
In this section we develop the PT in the effective expansion parameter
$\epsilon = \frac{1}{N^{1/3}}$ for the graphs BSDA in the frameworks
of the second Legendre transform. Consideration of this problem is
quite general and independent of the specific form of the
Lagrangian. So, it will be carried out in condensed notations of
Refs. \cite{Bordag1}-\cite{pl}.

The second Legendre transform is introduced by  representing  
the connected Green functions in the form
\begin{equation} \label{slt}
W = S[0] + \frac{1}{2} Tr log \beta - \frac{1}{2} \Delta^{-1} \beta +
W_2,
\end{equation}
where $S[0]$ is the tree level action, $\beta$ is the full propagator
of the scalar N-component field, $\Delta $ is the free field
propagator, $W_2$ is the contribution of two particle irreducible
(2PI) graphs taken out of the connected Green functions and having the
functions $\beta(p)$ on the lines. The symbol "Tr" means summation
over discrete Matsubara frequencies and integration over a three
momentum (see for more details Refs. \cite{Bordag1}-\cite{pl}).

The propagator is related to the mass operator by the Schwinger-Dyson
equation
\begin{eqnarray} \label{sde}
\beta^{-1}(p) = \Delta^{-1}(p) - \Sigma(p),\\
\Sigma(p) = 2 \frac{\delta W_2}{\delta \beta(p)}.~~~~~~~~~~ 
\end{eqnarray} 
The general expressions (\ref{slt}) and (\ref{sde}) will be the starting
point for the construction of the BSDA PT. Calculations in SDA have
been carried out already in \cite{prd}-\cite{pl} and delivered 
the masses of the fields and the characteristics of the phase transition:
$T_c$ - transition temperature, and $ T_+, T_-$ - upper and lower
spinodal temperatures. These parameters will be used in the new PT as
the input parameters (zeroth approximation). Contributions of BSDA
diagrams will be calculated perturbatively.

First let us write the propagator in the form
\begin{equation} \label{fp}
\beta(p) = \beta_0(p) + \beta^{'}(p),
\end{equation}
where  $\beta_0$ is derived in the SDA and
 $\beta{'}$ is a correction  which has to be calculated
in the BSDA PT in the small parameter $\epsilon = \frac{1}{N^{1/3}}$
for large $N$. The 2PI  part can be presented as 
\begin{equation} \label{w2}
W_2 = W_{SD} + W'_2 = W_{SD}[\beta_0 + \beta^{'}]  + W'_2[\beta_0 +
\beta^{'}]
\end{equation}
and assuming $\beta{'}$ to be small of order $\epsilon$ value we write
\begin{eqnarray} \label{expansion}
W_{SD} &=& W_{SD}[\beta_0] + \frac{\delta
W_{SD}[\beta_0]}{\delta\beta_0} \beta{'} + O(\epsilon^2),\\
  W'_{2} &=& W'_{2}[\beta_0] + \frac{\delta
W'_{2}[\beta_0]}{\delta\beta_0} \beta{'} + O(\epsilon^2).
\end{eqnarray}
In the above formulas the squared brackets denote a functional
dependence on the propagator. The curly brackets as usual mark a
parameter dependence.  In such a way other functions can be
expanded. For $\beta^{-1}$ we have
\begin{eqnarray} \label{bet}
\beta^{-1} &=& \beta_0^{-1} - \beta_0^{-1}\frac{\beta^{'}}{\beta_0}~~~~~~~~~~\\
~~~~~~~~~~~&=& \Delta^{-1} - \Sigma_0[\beta_0] - \Sigma^{'}[\beta_0],
\end{eqnarray}
where $\Sigma^{'}[\beta_0] = 2 \frac{\delta W_2[\beta_0]}{\delta
\beta_0}$ is a correction following from the 2PI graphs,
$\Sigma_0[\beta_0](p)$ is the super daisy mass operator. In a high
temperature limit within the ansatz adopted in
Refs.\cite{Bordag1}-\cite{pl} ($\beta^{-1} = p^2 + M^2$) it looks as
follows
\begin{equation} \label{be'}
\beta^{-1}(p) =  p^2 + M^2_0 - \Sigma^{'}[\beta_0](p),
\end{equation}
where $ M^2_0$ is the field mass calculated in the SDA as the solution
of the gap equations.

In a similar way, the free energy functional can be presented as
\begin{equation} \label{fe}
W = W^{(0)} + W'
\end{equation}
with
\begin{equation} \label{fe0}
 W^{(0)} = S^{(0)} + \frac{1}{2} Tr log \beta_0 - \frac{1}{2} Tr
 \beta_0 \Delta^{-1} + W_{SD} [\beta_0]
\end{equation}
and
\begin{equation} \label{fe1}
 W' = - \frac{1}{2} Tr \beta ' \Delta^{-1} + \frac{\delta
 W_{SD}[\beta_0] }{\delta \beta_0} \beta {'} + W_2^{'}[\beta_0]
 +\frac{1}{2} Tr \beta ' \beta^{-1}_0 + O(\epsilon^2).
\end{equation}
Taking into account that $\beta^{'} = 2 ~\beta_0^2 ~ \frac{\delta
W^{'}_{2}[\beta_0] }{\delta \beta_0}$ one finds
\begin{equation} \label{fe2}
\frac{1}{2} Tr \beta ' \beta^{-1}_0 = Tr \beta_0 \frac{\delta
W_{2}[\beta_0]' }{\delta \beta_0 },
\end{equation}
and hence
\begin{equation} \label{fe3}
W ' = W '_2[\beta_0].
\end{equation}

Thus, within the representation (\ref{fp}) we obtain for the $W$
functional
\begin{equation} \label{slt1}
W = W_{SD}^{(0)}[\beta_0] + W_2^{'}[\beta_0],
\end{equation}
where $ W_{SD}^{(0)}[\beta_0]$ is the expression (\ref{slt})
containing as the $W_2[\beta_0]$ the SDA part only and the particle
masses have to be calculated in the SDA. The term $W_2^{'}[\beta_0]$
corresponds to the 2PI graphs taken with the $\beta_0$ propagators on
lines.

From the above consideration it follows that perturbative calculations
in the parameter $\epsilon$ derived in the SDA within the second
Legendre transform can be implemented in a simple procedure including
as the input masses of propagators $\beta$ the ones obtained in the
SDA. Different types of the BSDA diagrams exist.  They can be
classified as the sets of diagrams having the same orders in
$\frac{1}{N}$ from the number of components. So, it is reasonable to
account for the contributions of the each class by summing up all
diagrams with the corresponding specific power of $\frac{1}{N}$. A
particular example of such type calculations will be done below.

\section*{3. Expansion near $T_c$}
In this section we shall calculate a first correction in the effective
expansion parameter $\epsilon = \frac{1}{N^{1/3}}$ for large $N$ at $T
\sim T_c$.

Before to elaborate that, let us take into consideration the main
results on the SDA which have to be used as a PT input. In
Ref. \cite{pl} it was shown that the masses of the Higgs $M_{\eta}$
and the Goldstone $M_{\phi}$ fields near the phase transition
temperature are ( for large N)
\begin{eqnarray} \label{mass}
M^{(0)}_{\eta} = \frac{\lambda T_+}{4 \pi}\Bigl( \frac{1}{(2N)^{1/3}}
- \frac{1}{2N}+ ...\Bigr) ,\\ \nonumber M^{(0)}_{\phi} = \frac{\lambda
T_+}{2 \pi} \Bigl( \frac{1}{(2N)^{2/3}} - \frac{1}{2N} + ...\Bigr) ,
\end{eqnarray}
where $\lambda$ is a coupling constant, upper script zero means that
in what follows these masses will be chosen as zero approximation, m -
initial mass in the Lagrangian. The upper spinodal temperature $T_+$
is close to the transition temperature $T_c \sim
\frac{m}{\sqrt{\lambda}}$ . We adduced here the masses for the $T_+$
case because that delivers simple analytic expressions. Results for
other temperatures in between, $T_- \le T \le T_+$, are too large to
be presented here. Note also that the temperatures $T_+, T_-$ in the
SDA are related as (see Refs.\cite{prd}, \cite{pl})
\begin{equation} \label{tt}
\frac{T_+}{T_-} = 1 + \frac{9 \lambda}{16 \pi^2} \frac{1}{(2N)^{2/3}} +
  ...,
\end{equation}
 and $T_- = \sqrt{\frac{12N}{\lambda (N + 2)}} m$. Hence it is clear
 that the transition is of a weakly first-order transforming into a
 second order one in the limit $N \to \infty$.

With these parameters taken into account an arbitrary graph beyond the
SDA having n vertexes can be presented as
\begin{equation} \label{vertex}
(\frac{\lambda}{N})^n T^C M^{3C-2L} V_n \sim (\frac{\lambda}{N})^n
(1/\sqrt{\lambda})^{C} ( \frac{\lambda}{N^{2/3}})^{3C-2L} V_n =
\lambda^1 (\frac{1}{N})^{\frac{1}{3}n + 2}V_n.
\end{equation}
Here the notation is introduced: C = L - n + 1 - number of loops, L -
number of lines. Since we are interested in diagrams with closed loops
only, the relation holds: 2n = L. The vertex factor $V_n$ comes from
combinatorics.  The right-hand-side of Eq.(\ref{vertex}) follows when
one shifts the three dimensional momentum of each loop, $\vec{p}
\rightarrow M \vec{p}$, and substitutes instead of M the mass
$M^{(0)}_{\phi}$ Eq.(\ref{mass}) to have the 'worst case' to
consider. In this estimate the static modes ($ l = 0$ Matsubara
frequency) of propagators were accounted for. One may wonder is it
sufficient at $T_c$? The positive answer immediately follows if one
takes into consideration that side by side with the three-momentum
rescaling the temperature component of the propagator is also shifted
as $T \to \frac{T}{M} \sim T N^{2/3}$. Hence it is clear that at $N$
goes to infinity a high temperature expansion is applicable and the
static mode limit is reasonable.

As it follows from Eq.(\ref{vertex}), $\lambda$ is not a good
expansion parameter near $T_c$ whereas $\frac{1}{N^{1/3}}$ is the one
because it enters in the power of the number of vertexes of the
graph. Of course, we have to consider the graphs beyond the SDA. That
is, all diagrams with closed loops of one line (tadpoles) have to be
excluded because they were summed up completely already at the SDA
level. Note also the important factor $\frac{1}{N^2}$ coming from the
rescaling of three-momentum with the mass $M_{\phi}^{(0)}$.

Before to proceed we note the most important advantages of
resummations in the SDA \cite{prd},\cite{pl}: 1) There is no an
imaginary part in the extrema of the free energy. 2) The simple ansatz
for the two-particle-irreducible Green function $\beta^{-1} = p^2 +
M^2$ is exact in this case. Here, $p$ is a four-momentum, M is a mass
parameter which is determined from the solution of gap equations. 3)
As it is known for many years, $T_c$ is well determined by this
approximation and it is not altered when further resummations are
achieved. 4) The existence of the expansion parameter $\epsilon =
\frac{1}{N^{1/3}}$ near $T_c$.

Since the estimate (\ref{vertex}) assumes the rescaling of momenta
$\vec{p} \rightarrow M \vec{p}$ the same procedure should be fulfilled
in perturbation calculations in the parameter $\epsilon$. They are
carried out in the following way. First, since $\lambda$ is not the
expansion parameter it can be skipped. Second, only the BSDA diagrams
have to be taken into consideration. Moreover, since $N$ is a large
number it is convenient to sum up series having different powers in
$\frac{1}{N}$. Third, rescaling $\vec{p} \rightarrow M\vec{p}$ can be
done before actual calculations. In this case one starts with the
expressions like in Eq.(\ref{vertex}) (for diagrams of $\phi$-sort to
consider). When $\eta$-particles are included one has to account for
the corresponding mass value and rescale the momentum
accordingly. Fourth, the temperature is fixed to be $T_c$. In other
words, one has to start with the expressions like in
Eq.(\ref{vertex}).

To demonstrate the procedure, let us calculate a series of graphs
giving leading in $\frac{1}{N}$ contribution $ F^{'}$ to the free energy
F. Then, the total to be $F = F^{(0)} + F^{'}$, where $F^{(0)}$ is the
result of the SDA.

The 'bubble chains' of $\phi$-field are the most divergent in N and
other diagrams have at least one power of N less. So, below we discuss
the $\phi$- bubble chains, only. The contribution of these sequences
with the mass $M_{\phi}^{(0)}$ and various n is given by the series
\begin{equation} \label{Df}
D_{\phi} = D_{\phi}^{(2)} + \frac{\lambda^1}{N^2} \sum_{n=3}^{\infty}
\frac{1}{n 2^n}\frac{V_n}{N^n}Tr_p (6 \Sigma^{(1)}_{\phi}(p) N^{2/3})^n.
\end{equation} 
Here, $D_{\phi}^{(2)}$ is the contribution of the "basketball"
diagram, $\Sigma^{(1)}_{\phi}(p)$ is the diagram of the type
\begin{equation} \label{b}
\Sigma^{(1)}_{\phi}(p) = Tr_k \beta^{(0)}_{\phi}(k) \beta^{(0)}_{\phi}(k+p) 
\end{equation}
and the power of the parameter $\lambda$ is written explicitly.

Now, let us sum up this series for large fixed N. The vertex
combinatorial factor for the diagram with n circles is \cite{prd}
\begin{equation} \label{Vn}
V_n =  \frac{N+3}{3} [(\frac{N+3}{3})^{n-1} + (\frac{2}{3})^{n-1} (N - 2)].
\end{equation}
The leading in N term in the $V_{n}$ is $\sim (N+1/3)^n$ and we have
for the series in Eq.(\ref{Df})
\begin{equation}
 \frac{\lambda^{1}}{N^2} \sum_{n=3}^{\infty}\frac{1}{n}Tr_p [(-
 \Sigma^{(1)}_{\phi}(p) N^{2/3})^n (N+1/N)^n].
\end{equation}
To sum up over n we add and subtract two terms: $-
\Sigma^{(1)}_{\phi}(p) N^{2/3}(N+1/N)$, and $\frac{1}{2}[-
\Sigma^{(1)}_{\phi}(p) N^{2/3} (N+1/N)]^2$.  We get
\begin{eqnarray}
D_{\phi} &=& D_{\phi}^{(2)} - \frac{\lambda^1}{N^2} Tr_p [log(1 +
\Sigma^{(1)}_{\phi}(p) N^{2/3}(N+1/N) ) \\ \nonumber &-&
\Sigma^{(1)}_{\phi}(p) N^{2/3}) (N+1/N) + \frac{1}{2}(-
\Sigma^{(1)}_{\phi}(p) N^{2/3} (N+1/N))^2].
\end{eqnarray}
To find all together we insert for the first term
\begin{equation}
D_{\phi}^{(2)} = \frac{1}{4} \lambda^1 \frac{(N^2 - 1)}{N^2}Tr_p
\frac{(- \Sigma^{(1)}_{\phi}(p))^2}{ N^{2/3}}.
\end{equation}
Then the limit N goes to infinity has to be calculated. We obtain
finally:
\begin{eqnarray} \label{F'}
D_{\phi} &=& - \frac{\lambda^1}{N^2}Tr_p log(1 +
\Sigma^{(1)}_{\phi}(p) N^{2/3})\\\nonumber &+&
\frac{\lambda^1}{N^{4/3}}Tr_p \Sigma^{(1)}_{\phi}(p) -
\frac{\lambda^1}{4 N^{2/3}}Tr_p (\Sigma^{(1)}_{\phi}(p))^2.
\end{eqnarray}
Thus, after summing over n the limit N goes to infinity exists and the
series is well convergent.  Eq.(\ref{F'}) gives the leading
contribution to the free energy calculated in the BSDA in the limit N
goes to infinity. This is the final result, if we are interested in
the leading in $\frac{1}{N}$ correction.

To obtain the squared mass, $M_{\phi}^2$, one has to sum up the value
calculated in the SDA $(M_{\phi}^{(0)})^2$ and - $\frac{2}{N - 1}
\delta D_{\phi}/\delta \beta_\phi(0)$, in accordance with a general
expression for the mass \cite{prd}. By substituting the propagators
one is able to find leading in $\frac{1}{N^{1/3}}$ corrections to free
energy and masses at temperatures close to $T_c$. In this way other
quantities can be computed.

The most important observation following from this example is that the
limit N goes to infinity does not commute with summing up of infinite
series in $n$ at $ n \rightarrow \infty$. It is seen that the first
two  terms in the Eq.(\ref{F'}) can be neglected as compared to the
last one which is leading in 1/N.  As it is occured, this contribution
is of order $N^{-5/3}$  that is smaller then the value of
$(M^{(0)}_{\phi})^2 \sim N^{- 4/3}$ (\ref{mass}) determined in  the
SDA. Note that
the correction is positive and the complete field mass is slightly
increased. We shall calculate the  value of the mass in sect.5
simultaneously with other parameters of the phase transition.
 Contributions of other next-next-to-leading classes of BSDA diagrams
 can also be obtained perturbatively.

\section*{4. Renormalization of  vortexes and  bubble chains}
Calculations carried out in the previous section deal with the
unrenormalized functions. But the question may arise: whether the
graphs, which are series of the $\Sigma^{1} (p, M_{\phi}) $ function
with the temperature dependent mass $M_{\phi}(T)$, are renormalized by
the temperature independent counter terms as one expects at finite
temperature? We shall consider in detail this question for the non
symmetric vertexes $V^{(n)}_{absd}(p)$ and the graphs $S^{(n)}(p)$ of
"bubble chain circles " with $n$ $\Sigma^{1} (p, M_{\phi}) $
insertions. These are of interest for computations in the previous
section. We will show that this is the case for both of these
objects. They contain, correspondingly, $n-1$ and $n$ insertions of
$\Sigma^{(1)} (p, M_{\phi}) $, $p = p_a + p_b$ is a momentum incoming
in the one-loop vertex $\Sigma^{1}_{[\beta_0]} (p, M_{\phi}) $. The
functions $S^{(n)}(p)$ are calculated from $V^{(n)}_{absd}(p)$ by
means of contracting the indexes and integrating over internal
momentum to form one extra $\Sigma^{1} (p, M_{\phi}) $ term. Note that
the Goldstone field bubble chains give a leading in $\frac{1}{N}$
contribution among the BSDA diagrams.
 
To carry out actual calculations we adopt the $O(N)$-model with the
notations introduced in Refs. \cite{prd}, \cite{pl}. In the restored
phase, we have $N$ scalar fields with the same masses $M_r(T).$ In the
broken phase, there is one Higgs field $\eta$ with the mass $M_{\eta}$
and $N - 1$ Goldstone fields $\phi$ having the mass $M_{\phi}$
(\ref{mass}) derived in the SDA.

Let us first consider the one-loop vertex of the $\phi$-field in the
s-channel, $s = p^2 = (p_a + p_b)^2$, $a, b$ mark incoming momenta,
\begin{equation} \label{v1}
V^{(2)}_{abc_1 d_1} = \frac{\rho^2 }{2}\frac{ \Sigma^1}{2} (C_1
s_{abc_1 d_1} + \frac{2}{3} V^1 _{abc_1 d_1}),
\end{equation}
where the notation is introduced: $\rho = \frac{-6 \lambda}{N}$ -
expansion parameter in the $O(N)$-model, $C_1 = \frac{N+1}{N-1} (
\frac{N+1}{3} - \frac{2}{3})$ is a combinatorial factor appearing at
the symmetric tensor $ s_{abc_1 d_1} = \frac{1}{3}\delta_{ab}
\delta_{c_1 d_1}$, $V^1 _{abc_1 d_1}= \frac{1}{3}(\delta_{ab}
\delta{c_1 d_1} + \delta_{a c_1} \delta_{b d_1} + \delta_{a d_1}
\delta_{b c_1})$ is the tree vertex in the $O(N)$-model. Subscript 1
in the indexes $c_1, d_1$ counts the number of loops in the
diagram. The diagram with $n$ loops ( $\Sigma^1_{[\beta_0]}(p)$
insertions) has the form
\begin{equation} \label{vn}
V^{(n+1)}_{abc_n d_n} = \frac{\rho^{(n+1)} }{2}(\frac{ \Sigma^1}{2})^n
(C_n s_{abc_n d_n} + (\frac{2}{3})^n V^1 _{abc_n d_n}),
\end{equation}
where now $C_n = \frac{N+1}{N-1} ( (\frac{N+1}{3})^n -
(\frac{2}{3})^n)$ and other notations are obvious.

At zero temperature, the function $\Sigma^1(p)$ has a logarithmic
divergence which in a dimensional regularization exhibits itself as a
pole $\sim \frac{1}{\epsilon}$, $\epsilon = d - 4$. We denote the
divergent part of $\Sigma^1(p)$ as $D_1$. To eliminate this part we
introduce into the Lagrangian the counter term $C_2$ of order
$\rho^2$:
\begin{equation} \label{C2}
C_2 = - \rho^2 \frac{D_1}{2} v^{(2)}_{abc_1 d_1},
\end{equation}
where $ v^{(2)}_{abc_1 d_1} = ( C_1 s + \frac{2}{3} V^1 )_{abc_1
d_1}$. Thus, the renormalized one-loop vertex is
\begin{equation} \label{v1r}
V^{(2)}_{abc_1 d_1} = \frac{1}{2} \rho^2 \frac{\Sigma^1_{ren.}}{2}
 v^{(2)}_{abc_1 d_1}
\end{equation}
with $\Sigma^{1}_{ren.} = \Sigma^1 - D_1$. 

In order $\rho^3$ three diagrams contribute. The first comes from the
tree vertexes $V^1_{abcd}$ and contains two $\Sigma^1$ insertions.
Two other graphs are generated by $V^1_{abcd}$ and the counter term
vertex $C_2$. Each of them has one $\Sigma^1$ insertion. The sum of
these three diagrams is
\begin{equation} \label{v3}
V^{(3)}_{abc_2 d_2} = \frac{1}{2} \rho^3 v^{(3)}_{abc_2 d_2}(
\frac{1}{4}( (\Sigma^1_{ren.})^2 + 2 \Sigma^1_{ren.}  D_1 + D_1^2 ) -
\frac{1}{2} D_1 \Sigma^1_{ren.} - \frac{1}{2} D_1^2 ).
\end{equation}
The terms with the products $\Sigma^1_{ren.}  D_1$ cancel in the
total. To have a finite expression one has to introduce a new counter
term vertex
\begin{equation} \label{c3}
C_3 = \frac{1}{4} \rho^3 D_1^2 v^{(3)}_{abc_2 d_2}.
\end{equation}
It cancels the last independent of $\Sigma^1$ divergence.  Thus the
renormalized vertex $V^{(3)}$ is given by the first term in the
expression (\ref{v3}).

This procedure can be easily continued with the result that in the
order $\rho^{(n+1)}$ one has to introduce the counter term of the
form
\begin{equation} \label{cn}
C_{n+1} = \rho^{n+1} (\frac{  D_1}{2})^n v^{(n+1)}_{abc_n d_n}.
\end{equation}
The finite vertex calculated with  all types of diagrams of the order
$\rho^{n+1}$ looks as follows
\begin{equation} \label{vn}
V^{n+1}_{abc_n d_n} = \frac{1}{2}\rho^{n+1} (\frac{
\Sigma^1_{ren.}}{2})^n v^{(n+1)}_{abc_n d_n}.
\end{equation}
Above  we considered the s-channel diagram contributions to the
vertex $V^{(n)}$. This is sufficient to study the leading in
$\frac{1}{N}$ correction $D_{\phi}$ (\ref{F'}) of interest. To have a
symmetric renormalized vertex one has to add the contributions of the
$t$- and $u$-channels and multiply the total by $\frac{1}{3}$.

Now it is easy to show that the counter terms $C_n$ rendering the
finiteness of the vertexes at zero temperature are sufficient to
renormalize them at finite temperature.  Really, $\Sigma^1(p, M(T),
T)$ can be divided in two parts, $\Sigma^1(p, M(T), T =0) =
\Sigma^1(p, M(T) ) $ corresponding to field theory and $\Sigma^1 (p,
M(T), T)$ - the statistical part. The divergent part $D_1$ and the
counter terms $C_n$ are independent of mass parameters. So, to obtain
the renormalized vertex at finite temperature $ V^{n+1}_{abc_n d_n}(p,
T)$ one has to sum up the same series of the usual and the counter
term diagrams as at zero temperature and substitute $\Sigma^1(p, M(T),
T) = \Sigma^1(p, M(T)) + \Sigma^1_T(p, M(T), T)$. Again, the terms of
the form $[\Sigma^1(p, M(T), T)]^{l} D_1^{n-l}$ are canceled and we
find the finite expression
\begin{equation} \label{vnt}
V^{n+1}_{abc_n d_n}(p, T, M(T)) = \frac{1}{2}\rho^{n+1} \Bigl(\frac{
  \Sigma^1_{ren.}(p, M(T)) + \Sigma^1_T(p, M(T), T) }{2} \Bigr)^n
  v^{(n+1)}_{abc_n d_n}.
\end{equation}
For this procedure to hold it is important that $D_1$ is
logarithmically divergent and independent of mass. So, the field
theoretical part of  $ V^{n+1}_{abc_n d_n}(p, T, M(T))$ as well as
the statistical part is renormalized by the same counter terms as the
vertex at zero temperature. In the BSDA PT we have to use the mass
$M^{(0)}_{\phi}(T)$.

Let us turn to the functions $S^{(n)}(p)$. They can be obtained from
the vertexes $ V^{n+1}_{abc_n d_n}(p, T, M(T))$ in the following
way. One has to contract the initial and the final indexes $a, b$ and
$c_n, d_n$ and form two propagators $\beta_0$ which after integration
over the internal line give the term $\Sigma^1$. The combinatorial
factor of this diagram is $\frac{1}{n + 1}$.

We proceed with the function $S^{(3)}$ at zero temperature.  In the
order $\rho^3$ two diagrams contribute. One includes three ordinary
vertexes $ V^1_{absd}$,
\begin{equation} \label{s31}
S^{(3)}_1(p) = 2 \frac{1}{3}
v^{(3)}_{abab}\Bigl(\frac{\Sigma^1(p)}{2}\Bigr)^3,
\end{equation}
and two other diagrams containing one vertex $v^1_{abcd}$ and the
counterterm vertex $C_2$,
\begin{equation} \label{s32}
2 S^{(3)}_2(p) = - \frac{1}{3}\rho^3 v^1_{abcd}
v^{(2)}_{cdab}\Bigl(\frac{\Sigma^1(p)}{2}\Bigr)^2 \frac{D_1}{2}.
\end{equation}
The contraction of indexes in Eq.(\ref{s32}) gives the factor
$v^{(3)}$ = $D_s^3 = v^{(3)}_{abab} = \frac{N +
1}{3}[(\frac{N+1}{3})^2 + (\frac{2}{3})^2 (N-2)]$. Again, substituting
$\Sigma^1 = \Sigma^1_{ren.} + D_1$ one can see that the terms with
products $(\Sigma^1_{ren.})^l (D_1)^{(3-l)}$ are canceled in the sum
of expressions (\ref{s31}) and (\ref{s32}). To obtain a finite
$S^{(3)}$ we introduce the counter term of order $\rho^3$,
\begin{equation} \label{C3}
C^{(3)} = \frac{1}{3}\rho^3 \Bigl(\frac{D_1}{2}\Bigr)^3 D_s^3.
\end{equation}
After that the renormalized circle $S^{(3)}$ is
\begin{equation} \label{s3r}
S^{(3)}_{ren.}(p) = \frac{1}{3}\rho^3
\Bigl(\frac{\Sigma^1_{ren.}(p)}{2}\Bigr)^3 D_s^3.
\end{equation}
Note that since $C^{(3)}$ does not depend on any parameter, it can be
omitted as well as the divergent term in the expression $S^{(3)}$. In
other words, there is no need to introduce new counter terms into the
Lagrangian in order to have a finite $S^{(3)}$ and the counter term
renormalizing the vertex $V^{(3)}$ are sufficient.

This procedure can be continued step by step for diagrams with any
number of $\Sigma^1$ insertions. The renormalized graph $ S^{(n)}$
looks as follows,
\begin{equation} \label{snr}
S^{(n)}_{ren.}(p) = \frac{1}{n}\rho^n
\Bigl(\frac{\Sigma^1_{ren.}(p)}{2}\Bigr)^n D_s^n,
\end{equation}
where the factor coming from the contraction of the
$v^{(n)}_{abc_{n-1} d_{n-1}}$ is
\begin{equation} \label{dn}
D^{n}_{s} = \frac{N+1}{3}\Bigl[ \Bigl(\frac{N+1}{3}\Bigr)^{n-1} +
\Bigl(\frac{2}{3}\Bigr)^{n-1} (N -2)\Bigr].
\end{equation}

Now, it is a simple task to show that the counter terms renormalizing
vertexes $V^{(n)}$ and circles $S^{(n)}$ at zero temperature are
sufficient to obtain finite $S^{(n)}(p, T)$ when the temperature is
switched on. This is based on the property that the temperature
dependent graph $\Sigma^1(p, M, T)$ can be presented as the sum of the
zero temperature part and the statistical part independently of the
mass term entering.  Then it is easy to check that the divergent terms
of the form $[\Sigma^1(p, M, T = 0) + \Sigma^1(p, M, T) ]^{l}
(D_1)^{n-l}$ are canceled when all the diagrams of order $\rho^n$
forming the circle $S^{(n)}_{ren.}(p, T)$ are summed up.

Thus we have shown, the renormalization counter terms of leading in
$1/N$ BSDA graphs calculated at zero temperature being independent of
the mass parameter entering $\Sigma^1$ renormalize the $S^{(n)}(p, T)$
functions at finite temperature. This gives the possibility to
construct a PT based on the solutions of the gap equations.  In fact,
just the series of $\Sigma^1$ functions are of interest at the
transition temperature $T_c$ which has to be considered as a fixed
given number. Other parameters of the phase transitions can be found
by means of some iteration procedure of the investigated already gap
equations.

\section*{5. Corrections to the parameters of the phase transition}

Having obtained the leading correction to the free energy (\ref{F'})
one is able to find perturbatively the characteristics of the phase
transition near $T_c$. We shall do that for the limit $N \to \infty$.

First let us calculate the corrections to the Higgs boson mass,
$\Delta M_{\eta}$, and the Goldstone boson mass, $\Delta M_{\phi}$,
due to $D_{\phi}$ term (\ref{F'}). The starting point of this
calculations is the system of gap equations derived in Ref.\cite{prd}
(Eqs. (28), (30) and (45)) and Ref. \cite{pl} (Eq. (28)).  Here we
write that in the form when only the term containing $D_{\phi}$ is
included,
\begin{eqnarray} \label{ge}
\frac{M^2_{\eta}}{2} &=&  m^2 - \frac{3 \lambda}{N} \Sigma^{(0)}_{\eta} 
- \lambda \frac{N - 1}{N}  \Sigma^{(0)}_{\phi},\\ \nonumber
\frac{M^2_{\phi}}{2} &=& \frac{ \lambda}{N} (\Sigma^{(0)}_{\phi} -
\Sigma^{(0)}_{\eta} ) - \frac{1}{N - 1} \frac{\delta D_{\phi}}{\delta 
\beta_{\phi}(0)}.
\end{eqnarray}
Remind that these equations give the spectrum of mass in the extrema
 of the free energy in the phase with broken symmetry. The complete
 system (Eqs.(45) in Ref.\cite{prd}) contains other terms which have
 higher orders in $\frac{1}{N}$ and were omitted. Here $m$ is the mass
 parameter in the Lagrangian. The functions $\Sigma^{(0)}_{\eta},
 \Sigma^{(0)}_{\phi}$ are the tadpole graphs with the full propagators
 $\beta_{\eta}, \beta_{\phi}$ on the lines. For the ansatz adopted in
 Refs.\cite{prd}, \cite{pl}, $\beta^{-1}_{\eta,\phi} = p^2 +
 M^2_{\eta,\phi}$, they have at high temperature the asymptotic
 expansion
\begin{equation} \label{s0}
\Sigma^{(0)}_{\eta, \phi} = Tr \beta_{\eta,\phi} = \frac{T^2}{12} -
\frac{M_{\eta,\phi} T}{4 \pi} + ... ,
\end{equation}
where dots mark next-next-to-leading terms. Within Eqs.(\ref{ge}) -
(\ref{s0}) (without the $D_{\phi}$ term) the masses (\ref{mass}) have
been derived in the limit of large $N$.

Now we compute the last term in the Eq.(\ref{ge}) for large $N$. In
this case the last term of $D_{\phi}$ in the Eq.(\ref{F'}) is dominant
and calculating the functional derivative we find
\begin{equation} \label{der}
f(M^{(0)}_{\phi}) = - \frac{2}{N - 1}~ \frac{\delta D_{\phi}}{\delta 
\beta_{\phi}(0)} = \frac{\lambda}{N - 1}~ \frac{1}{N^{2/3}}~ Tr \beta^3
(M^{(0)}_{\phi}).
\end{equation}
The sunset diagram entering the right-hand-side can be easily computed
to give \cite{prd}
\begin{equation} \label{der1}
 Tr \beta^3 (M^{(0)}_{\phi}) = \frac{T^2}{32 \pi^2}\Bigl ( 1 - 2~ ln
 \frac{3 M^{(0)}_{\phi}}{m}\Bigr) .
\end{equation}
Thus for $f(M^{(0)}_{\phi})$ we obtain in the large $N$ limit
\begin{equation} \label{f}
 f(M^{(0)}_{\phi})  = \frac{\lambda}{N^{5/3}}~ \frac{T^2}{32
 \pi^2}\Bigl ( 1 - 2~ ln \frac{3 \sqrt{3\lambda} }{\pi (2N)^{2/3}}\Bigr),
\end{equation}
where the mass $M^{(0)}_{\phi}$ (\ref{mass}) was inserted and $T =
T_+$ has to be substituted. Here we again turn to the $T_+$ case to
display analytic results.

Since $ f(M^{(0)}_{\phi})$ is small, it can be treated perturbatively
when the masses $M_{\eta}$ and $ M_{\phi}$ are calculated. Let as
write them as
\begin{eqnarray} \label{mass1}
M_{\eta} &=& M_{\eta}^{(0)} + x, \\ \nonumber
M_{\phi} &=& M_{\phi}^{(0)} + y 
\end{eqnarray}
assuming $x, y$ to be small. Substituting these in the  equations
(\ref{ge}) - (\ref{s0}) and preserving linear in $x, y$ terms we
obtain the system 
\begin{eqnarray} \label{ge1}
2 M_{\phi}^{(0)}~ y &=& \frac{\lambda T}{2 \pi N}~ (x - y) +
f(M_{\phi}^{(0)}),\\ \nonumber
2 M_{\eta}^{(0)}~ x &=& \frac{3 \lambda T}{2 \pi N}~ x + \frac{\lambda (N
  - 1) T}{2 \pi N}~ y
\end{eqnarray}
of linear equations. Its solutions for large $N$ are
\begin{eqnarray} \label{mass2}
~~~~~~x &=& \frac{1}{3}~ \frac{T_+}{32 \pi}~\frac{1}{ N^{2/3}}\Bigl (
1 - 2~ ln \frac{3 \sqrt{3\lambda} }{\pi (2N)^{2/3}}\Bigr) , \\
\nonumber ~~~~~~y &=& \frac{1}{2}~ \frac{2^{2/3}}{3^{1/2}}~
\frac{T_+}{32 \pi}~\frac{1}{ N}\Bigl ( 1 - 2~ ln \frac{3
\sqrt{3\lambda} }{\pi (2N)^{2/3}}\Bigr) .
\end{eqnarray}
As one can see, these corrections are positive numbers smaller than
the masses (\ref{mass}) calculated in the SDA, as it should be in a
consistent PT. Notice that the correction $x$ is larger as compared to
the next to leading term in the SDA $\sim (1/N)$.  So, the BSDA graphs
deliver the next-to-leading correction. The value $y$ is of the same
order as the next-to-leading term in the SDA.

In a similar way the correction to the mass in the restored phase,
$\Delta M_r$, can be calculated. In this symmetric case the all
components have equal masses which in the SDA
within the representation (\ref{s0}) are the solutions of the gap
equation (Eq. (36) in Ref.\cite{prd} and Eq.(27) in Ref. \cite{pl})
\begin{equation} \label{ger}
M_r^2 =  - m^2 + \frac{\lambda (N + 2) T^2}{12 N} - 2 M_r \frac{\lambda
  (N + 2) T}{8 \pi N}.
\end{equation}
It has a simple analytic solution
\begin{equation} \label{mr}
M_r^{(0)} = - \frac{\lambda (N + 2)T}{8 \pi N} + \sqrt{\Bigl(\frac{\lambda
    (N + 2)T}{8 \pi N}\Bigr)^2 - m^2 + \frac{\lambda (N + 2)T^2}{12 N}},
\end{equation}
where again the upper script zero means the SDA result. The
contribution of the bubble graphs (\ref{F'}) corresponds either to the
broken or to the restored phases. The only difference is in the number
of contributing field components. In the broken phase this is $N -1$
and in the restored it is $N$. But it does not matter for large $N$.
So, the function which has to be substituted into Eq. (\ref{ger}) is
the one in Eq. (\ref{der}) with the replacements $N - 1 \to N$ and
$M^{(0)}_{\phi} \to M^{(0)}_{r}$.

To calculate the mass correction $z$ which is assumed to be small we
put $M_r = M^{(0)}_{r} + z $ into Eq.(\ref{ger}) and find in the limit
$N \to \infty$:
\begin{equation} \label{z}
~~~~~z = \frac{ f(M^{(0)}_{r})}{2 M^{(0)}_{r} + \frac{\lambda T}{ 4
      \pi} }.
\end{equation}
We see that the correction has a small positive value. By requiring
that $M_r$ to be positive the temperature $T_-$ can be simply
calculated. It is clear that $T_- \le T_-^{(0)}$ and BSDA graphs
decrease slightly the lower spinodal temperature.

Now we calculate the correction to $T_+$. To do that let us turn again
to the system (\ref{ge}) with the expressions (\ref{s0}), (\ref{f})
been substituted. From the second equation of the system we find
\begin{equation} \label{me}
M_{\eta} = M_{\phi} + \frac{2 \pi N}{\lambda T} (M_{\phi}^2 -
  f(M_{\phi}^{(0)}))
\end{equation}
and insert this into the first equation to have
\begin{eqnarray} \label{emf}
\Bigl(\frac{2\pi N}{\lambda T}\Bigr)^2 ( M_{\phi}^4 - 2 M_{\phi}^2 f )
+ M_{\phi}^2 + \frac{4\pi N}{\lambda T} (M_{\phi}^3 - M_{\phi} f )
=~~~~~~~~~~~~~~~~~~~~~ \\ \nonumber 2 m^2 - \frac{\lambda (N + 2)
T^2}{6 N} + \frac{3 \lambda T}{2 \pi N} \Bigl( M_{\phi} + \frac{2 \pi
N}{\lambda T} ( M_{\phi}^2 - f)\Bigr) + \frac{\lambda (N - 1) T}{2 \pi
N } M_{\phi},
\end{eqnarray}
where the linear in $f$ terms are retained. This fourth order
algebraic equation can be rewritten in the dimensionless variables
$\mu = \frac{2 \pi N}{\lambda T} M_{\phi}$ and $g = (\frac{2 \pi
N}{\lambda T})^2 f$ as follows:
\begin{eqnarray} \label{emf1}
F(\mu) &=& 0,~~~~~~~~~~~~~~~~~~~~~~~~~~~~~~~~~~~~~~~~~~~~~~~~~~~~~~\\
\nonumber F(\mu) &=& \mu^4 + 2\mu^3 - 2 \mu^2 - h - ( N + 2) \mu - 2 g
( \mu^2 + \mu ).
\end{eqnarray}
Here the $\mu$-independent function is
\begin{equation} \label{h}
h = h^{(0)} - 3 g = (2 m^2 - \frac{\lambda (N + 2) T^2}{6 N} )(\frac{2\pi
  N}{\lambda T})^2 - 3 g
\end{equation}
and as before the SDA part is marked by the upper script zero. Remind
that this equation determines the Goldstone field masses in the
extrema of the free energy. It has two real solutions corresponding to
a minimum and a maximum. The equation (\ref{emf1}) has roots for any
$N$ and $T$ which are too large to be displayed here. To have simple
analytic results and investigate the limit of large $N$ we proceed as
in Ref.\cite{pl} and consider the condition for the upper spinodal
temperature. In the case of $T = T_+$ the two solutions merge and we
have a second equation,
\begin{equation} \label{f'}
F^{'}(\mu) = \mu^3 + \frac{3}{2} \mu^2 - \frac{1}{2} \mu -
(\frac{N}{4} + \frac{1}{2}) - g( \mu + \frac{1}{2}) = 0. 
\end{equation}
It can be easily solved for large $N$: 
\begin{equation} \label{sol}
\mu = \Bigl(\frac{N}{4}\Bigr)^{1/3} - \frac{1}{2} + ...~.
\end{equation} 
Here the first two terms of asymptotic expansion are included that is
sufficient for what follows. Notice that since $g$-dependent terms are
of next-next-to-leading order they do not affect the main contributions.
The function $h$ can be calculated from Eq.(\ref{emf1}) and its first
two terms are
\begin{equation} \label{s0lh}
 h^{(0)} = - 3 \Bigl(\frac{N}{4}\Bigr)^{4/3} - \frac{7}{2} \Bigl( \frac{N}{4}\Bigr)^{2/3}  + ...~.
\end{equation}
The temperature $T_+$ computed from Eq.(\ref{h}) can be rewritten in
the form
\begin{equation} \label{t+}
T_+ = \sqrt{2} \frac{2\pi N}{\lambda} \Bigl( \frac{2 \pi^2 (N + 2)
  N}{3 \lambda } + h^{(0)} - 3 g \Bigr)^{- 1/2},
\end{equation}
where the values of $h^{(0)}$ Eq.(\ref{s0lh} ) and $g$ Eq. (\ref{f})
have to be substituted. Again, since $g$ is a small
next-next-to-leading term the solution can be obtained
perturbatively. We find
\begin{equation} \label{t+1}
T_+ = T_+^{(0)} \Bigl( 1 + \frac{9 \lambda}{32 \pi^2} \frac{ ( 1 - 2~
ln \frac{3 \sqrt{3\lambda} }{\pi (2N)^{2/3}})}{N^{5/3}}\Bigr) ,
\end{equation}
where the value $T_+^{(0)} = T_-^{(0)}( 1 + \frac{9 \lambda}{16 \pi^2}
\frac{1}{(2N)^{2/3}} )$ must be inserted. From this result it follows
that the upper spinodal temperature is slightly increased due to the
BSDA contributions. But this is next-next-to-leading correction to the
$T_+^{(0)}$ of the SDA.

Thus, we have calculated the main BSDA corrections to the particle
masses and the upper and lower spinodal temperatures in the
$(\frac{1}{N})^{1/3}$ PT. We found that $T_-$ is decreased and $T_+$
is increased as compared to the SDA results due to the leading in
$\frac{1}{N}$ BSDA graphs - bubble chains. So, the strength of the
first-order phase transition is slightly increased when this
contribution is accounted for. However, this is a next-to-leading
effect. In such a way we prove the results on the type of the phase
transition obtained already in the SDA \cite{prd}, \cite{pl}.

\section*{6. Discussion}

As the carried out calculations showed, the phase transition in the
$O(N)$-model at large finite $N$ is weakly of first-order. It becomes
a second order one in the limit $N \to \infty$. This conclusion has
been obtained in the SDA and was proved in the perturbative
calculations in the consistent BSDA PT in the effective expansion
parameter $\epsilon = (\frac{1}{N})^{1/3} $. This parameter appears
near the phase transition temperature $T_c$ at the SDA level. Let us
summarize the main steps of the computation procedure applied and the
approximations used to derive that.

In Refs.\cite{Bordag1} - \cite{pl} as a new method the combination of
the second Legendre transform with considering of gap equations in the
extrema of a free energy functional was proposed. This has simplified
calculations considerably and resulted in transparent formulas for
many interesting quantities.  Within this approach the phase
transitions in the $O(1)$ - and $O(N)$-models were investigated in the
super daisy and beyond approximations. To have analytic expressions a
high temperature expansion was systematically applied and the ansatz
for the full propagators $\beta^{- 1} = p^2 + M^2$ has been used. This
ansatz is exact for the SDA which sums up completely the tadpole
graphs.  Just within these assumptions a first order phase transition
was observed and the effective parameter $\epsilon$ has been found. In
terms of it all interesting characteristics can be expressed in the
limit of large $N$ and perturbation theory at $T \sim T_c$
constructed. This solves an old problem on the choice of a zero
approximation for perturbative computations near $T_c$ (if such
exist).  We have shown explicitly here that these are the SDA
parameters taken as the input values for BSDA resummations. Clearly,
this does not change qualitatively the results obtained in the SDA.

Two important questions should be answered  in connection with the
  results presented. The first is on the relation with other investigations
where a second order phase transition was observed (see, for example,
 the well known literature \cite {Zinn} - \cite{Fodor}), and in fact this is a
general opinion. The second is the non zero mass
(\ref{mass}) of the Goldstone excitations in the broken phase  at
finite temperature  that was determined in SDA by solving of gap
equations \cite{prd}, \cite{pl}. 

What concerns the first question, we are able to analyze the papers
\cite{Tetradis}, \cite{Reuter} \cite{Elmfors},
\cite{Ogure1},\cite{Ogure2}, where analytic calculations have been
reported. Partially that was done in Refs.\cite{prd}, \cite{pl}. We
repeat that here for completeness.  First of all we notice that there
are no discrepancies for the limit $N \to \infty$ where all the
methods determine a second order phase transition. The discrepancy is
for large finite $N$.

In Ref.\cite{Elmfors} the renormalization group method at finite
temperature has been used and a second order phase transition was
observed. Results obtained in this approach are difficult to compare
with that found in case of a standard renormalization at zero
temperature. This is because a renormalization at finite temperature
replaces some resummations of diagrams which remain unspecified
basically. In Refs.\cite{Ogure1}, \cite{Ogure2} a non-perturbative
method of calculation of the effective potential at finite temperature
- an auxiliary field method - has been developed and a second order
phase transition was observed in both, the one- and the $N$-component
models. This approach seems to us not self-consistent because it
delivers an imaginary part to the effective potential in its
minima. This important point is crucial for any calculation procedure
as a whole. Really, the minima of an effective potential describe
physical vacua of a system. An imaginary part is signaling either the
false vacuum or the inconsistency of a calculation procedure
used. This is well known beginning from the pioneer work by Dolan and
Jackiw \cite{Dolan} who noted the necessity of resummations in order
to have a real effective potential. In our method of calculation this
requirement is automatically satisfied when the SDA diagrams are
resummed in the extrema of free energy. This consistent approximation
is widely used and discussed in different aspects in literature
nowadays \cite{Drummond1}-\cite{Peshier}.

Let us discuss in more detail the results of Refs.\cite{Tetradis},
\cite{Reuter} where an interesting method - the method of average
potential - was developed and a second order phase transition has been
observed for any value of $N$. To be as transparent as possible we
consider the equation for the effective potential derived in
Ref.\cite{Reuter} (equation (3.13))
\begin{equation} \label{u'}
U^{'}(\rho, T) = \bar{\lambda} \Bigl( a^2 + \rho - b~
  \sqrt{U^{'} (\rho, T)} \Bigr),
\end{equation}
where the notation is introduced: $U^{'} (\rho, T)$ is the derivative
with respect to $\rho$ of $U (\rho, T)$,~ $\rho = \frac{1}{2} \phi^a
\phi^a$ is the condensate value of the scalar field, $\bar{\lambda}$
is the coupling constant. The parameters $a^2$ and $b$ are:
\begin{eqnarray} \label{ab}
a^2 &=& ( T^2 - T_{cr}^2) \frac{N}{24},~~ T_{cr}^2 = \frac{24
  m^2_R}{\lambda_R N}, \\ \nonumber
b &=& \frac{NT}{8 \pi}.~~~~~~~~~~~~~~~~~~~~~~~~~~~~~~~
\end{eqnarray}
Here $ m_R$ and $\lambda_R$ are renormalized values, and
$\lambda_R = \bar{\lambda} (1 + \frac{\bar{\lambda}N}{32 \pi^2} ln
(L^2/M^2))^{-1}$ (see for details the cited paper).

Considering the temperatures $T \sim T_{cr}$ and $\rho << T^2$ the
authors have neglected $U^{'}(\rho, T)$ as compared to the
$\sqrt{U^{'} (\rho, T)}$ in Eq.(\ref{u'}) and obtained after
integration over $\rho$ (formula (3.16) of Ref. \cite{Reuter})
\begin{equation} \label{u}
U (\rho, T) = \frac{\pi^2}{9} \Bigl(\frac{T^2 - T_{cr}^2}{T}\Bigr)^2
\rho + \frac{1}{N} \frac{8 \pi^2}{3} \frac{T^2 - T_{cr}^2}{T^2} \rho^2
+ \frac{1}{N^2} \frac{64 \pi^2}{3}\frac{1}{T^2}\rho^3. 
\end{equation}
Near $T \sim  T_{cr}$ this effective potential includes  the  $\rho^3
(\phi^6)$ term and predicts a second order phase transition. That is
the main conclusion.  

However, this analysis is insufficient to distinguish between a second
and a weakly-first-order phase transitions.  Actually, the temperature
$T_{cr}$ determined from the initial condition $U'(0, T_{cr}) = 0$ in
the former case corresponds to the temperature $T_-$ in the latter
one. To determine the type of the phase transition one has to verify
whether or not a maximum at finite distances from the origin in the
$\rho$-plane exists. To check that the equation (\ref{u'}) has to be
integrated exactly, without truncations. This is not a difficult
problem if one first solves Eq.(\ref{u'}) with respect to $U^{'}
(\rho, T)$,
\begin{equation} \label{u'1}
U^{'} (\rho, T) = \bar{\lambda} \Bigl( a^2 + \frac{\lambda b^2}{2} +
\rho - b~ \sqrt{\bar{\lambda} (a^2 + \rho) + \frac{\bar{\lambda}^2
    b^2}{4} } \Bigr),
\end{equation}
and then integrates to obtain
\begin{eqnarray} \label{u1}
U (\rho, T) &=& \Bigl( \bar{\lambda} (T^2 - T_{cr}^2)\frac{N}{24} +
\frac{\bar{\lambda}^2 T^2 N^2}{128 \pi^2} \Bigr) \rho + \frac{1}{2}
\rho^2 \\ \nonumber
&-& \frac{2}{3} \frac{\bar{\lambda} T N}{8 \pi} \Bigl( (T^2 -
T_{cr}^2)\frac{N}{24} +\frac{\bar{\lambda} T^2 N^2}{256 \pi^2} +
 \rho \Bigr)^{3/2},
\end{eqnarray}
where the values of $a$ and $b$ Eq.(\ref{ab}) are accounted for. We
see that in contrast to Eq.(\ref{u}) this potential includes a cubic
term which is responsible for the appearance of a maximum in some
temperature interval and a first order phase transition, that is quite
known. Obviously, expanding the expression (\ref{u1}) at $\rho \to 0 $
and retaining three first terms, one reproduces the $\rho^3$ term of
Eq.(\ref{u}).

This consideration convinces us that there are no discrepancies with
the actual results of the average action method. The expression
(\ref{u'}) gives the potential (\ref{u1}) predicting a first order
phase transition.

Our final remarks are on the Goldstone theorem at finite temperature.
In fact, at $T \not = 0$ the Goldstone bosons should not inevitably be
massless as it was argued, in particular, in
Ref.\cite{Kowalski}. Formally, the reason is that at finite
temperature Lorentz invariance is broken and therefore the condition
$p^2 = 0$ does not mean zero mass of the particle in contrast to as it
should be at $T = 0$ (see for details Ref.\cite{Kowalski}).  We
observed in the SDA that this is the case when the first order phase
transition happens. In the limit $N \to \infty$ corresponding to a
second order phase transition the Goldstone bosons are massless, as it
is seen from Eq.(\ref{mass}). The same conclusion follows from the
results of Ref.\cite{Reuter} when a second order phase transition is
assumed. Probably this problem requires a separate investigation by
means of other methods.

One of the authors (V.S.) thanks Institute for Theoretical Physics
University of Leipzig for hospitality when the final part of this work
has been done.

\end{document}